\documentstyle[prb,aps,epsf]{revtex}

\begin{document}

\title{
Quantum Measurements Performed with a Single-Electron Transistor}

\author {Alexander Shnirman and Gerd Sch\"on}

\address{
Institut f\"ur Theoretische Festk\"orperphysik,
Universit\"at Karlsruhe, D-76128 Karlsruhe, Germany}

\maketitle

\begin{abstract}
Low-capacitance Josephson junction systems as well as coupled 
quantum dots, in 
a parameter range where single charges can be controlled,
provide physical realizations of quantum bits, discussed in
connection with quantum computing. 
The necessary manipulation of the quantum states
can be controlled by applied gate voltages.
In addition, the state of the system has to be read out.
Here we suggest to measure the quantum state by coupling 
a single-electron transistor to the q-bit.  
As long as no transport voltage is 
applied, the transistor 
influences the quantum dynamics of the q-bit only weakly. We have
analyzed the time evolution of the density matrix of the 
transistor and q-bit when a voltage is turned on.
For values of the capacitances
and temperatures which can be realized by modern nano-techniques the
process constitutes a quantum measurement process.\\
\end{abstract}

\section{Introduction}
Recent proposals \cite{Our_PRL,Mooij,Averin,Loss} suggested to 
use nanoscale devices, such as low-capacitance Josephson junctions
or coupled quantum dots as quantum bits (q-bits), which are the basic
elements of quantum computers. The two logical states are different
charge states of the system \cite{Our_PRL,Mooij,Averin}. 
Applied gate voltages allow the necessary controlled manipulations 
(single-bit and two-bit operations) of the quantum states. 
In addition to these manipulations, a read-out device is required
to perform quantum measurements of the resulting state of the
q-bit. We suggest to use single-electron transistors for this purpose.

The requirements to perform, on one hand, quantum manipulations 
and, on the other hand, a quantum measurement appear to contradict 
each other. During the manipulations the dephasing should
be minimized, while a quantum measurement should dephase 
the state of the q-bit as fast as possible. The option
to couple the measuring device to the
q-bit only when needed is hard to achieve
in mesoscopic systems.  The alternative, which we discuss here, is to
keep the measuring device permanently coupled to the q-bit in a state of
equilibrium during the quantum operations. The measurement 
is performed by driving the measuring device out of equilibrium,
in a way which dephases the quantum state of the q-bit. 
Similar nonequilibrium dephasing processes have recently been considered
by a number of authors \cite{Levinson,Aleiner,Gurvitz,Buks}.

For definiteness we discuss in this paper the measurement process 
performed by a single-electron tunneling (SET) transistor coupled 
capacitively to a Josephson junction q-bit; however,
this type of measurements may be performed for any quantum system 
with two different charge states.
We describe the measuring process by 
considering the time-evolution of the density matrix of the coupled
system. We show that the process is characterized by three different 
time scales: the dephasing time, the time of measurement, which may be 
longer than the dephasing time, and the mixing time, i.\,e. the time
after which all the information about the initial quantum state is lost
due to the transitions induced by the measurement.  
Thus, we arrive at a new criterion for a ``good'' quantum measurement:
the mixing time should be longer than the time of measurement.

\section{The quantum system and the measuring device}
The system is shown in Fig.~\ref{CIRCUIT}.
The two superconducting islands in the upper part
are the realization of a q-bit. Its state 
is characterized by a discrete variable, $n$, 
the number of extra Cooper 
pairs on the lower superconducting island. 
The lower part (a normal island between two normal leads) 
stands for a SET transistor, which is coupled capacitively 
to the q-bit. It's charging state is characterized by the 
extra charge on the middle island, $eN$.
A similar setup has recently been studied in the 
experiments of Refs.~[\onlinecite{Bouchiat,Bouchiat_PhD}]
with the purpose to demonstrate that the ground
state of a single Cooper pair box is a coherent 
superposition of different charge states. We discuss the 
relation of our proposal to these experiments below.

As shown earlier \cite{Our_PRL} the quantum operations 
with the q-bit are performed by controlling the applied 
gate voltage $V_{\rm qb}$.
At this stage the transport voltage $V$ across  the SET
transistor is kept zero.
Therefore no dissipative currents flow in the system, and 
dephasing effects due to the transistor are minimized.
To perform a measurement one applies a transport voltage $V$.
The resulting normal current through the transistor
depends on the charge  configuration of the q-bit,
since different charge states induce different
voltages on the middle island of the SET transistor. 
In order to check whether the dissipative
current through the SET transistor contains information about
the quantum state of the q-bit, we have to discuss various 
noise factors (shot noise) and  the  measurement 
induced transitions between the states of the q-bit.
It turns out that for suitable parameters, which can be realized
experimentally, the dephasing by the passive SET transistor is
weak. When the transport voltage is turned on the dephasing is fast,
and the current through the transistor --  after a transient period --
provides a measure of the state of the q-bit. At still longer times the
complicated dynamics of the composite system destroys the information
of the quantum state to be measured.

The Hamiltonian of the composite system consists of three main parts:
the charging energy, the terms describing the microscopic
degrees of freedom of the metal islands and electrodes, 
and the tunneling terms, including the Josephson
coupling. The charging term is a quadratic form in the 
variables $n$ and $N$:
\begin{eqnarray}
\label{CHARGING_ENERGY}
H_{\rm charge}&=&E_{\rm qb} n^2 + E_{\rm set} N^2 + E_{\rm int} nN +
\nonumber \\
&&2en V_{n} + eN V_{N} + {\rm const.} \ .
\end{eqnarray} 
The charging energy scales $E_{\rm qb}$, $E_{\rm set}$ and $E_{\rm int}$ are 
determined by the capacitances between all the islands. Similarly,
the effective gate voltages  $V_{n}$ and $V_{N}$  
depend in general on all three voltages $V_{\rm qb}$, $V_g$ and $V$, 
but for a symmetric bias (see Fig.~\ref{CIRCUIT}),  
$V_{n}$ and $V_{N}$ are controlled only by
the two gate voltages, $V_{\rm qb}$ and $V_g$ .

The microscopic terms $H_{\rm L}$, $H_{\rm R}$ and $H_{\rm I}$ describe 
noninteracting electrons in the two leads and on the middle island 
of the SET transistor, respectively:
\begin{equation}
\label{FREE_LEADS_H}
        H_r =  \sum_{k\sigma} \epsilon_{k\sigma}^{r} c_{k\sigma}^{r\dag} 
        c_{k\sigma}^{r \phantom{\dag}}
        \ \ (r= {\rm L,R,I})\ , 
\end{equation}
The index $\sigma$ labels the transverse channels including the spin,
while $k$ labels the wave vector within one channel. 
Similar terms exist for the two islands
of the q-bit. Here we use the ``macroscopic'' description
of the superconductors, assuming that the microscopic degrees of
freedom have already been integrated out \cite{Schoen_Zaikin}.

The tunneling terms include the Josephson coupling
$H_{\rm J} = - E_{\rm J}\cos\Theta$, which describes the transfer 
of Cooper pairs between the two islands of the q-bit 
($e^{i\Theta} |n\rangle = |n+1\rangle$), 
and the normal tunneling Hamiltonian for the SET transistor: 
\begin{equation}
\label{TUNNEL_HAMILTONIAN}
        H_{\rm T} = \sum\limits_{kk'\sigma} 
        T^{L}_{kk'\sigma} c^{\rm L\dag}_{k\sigma} 
        c_{k'\sigma}^{\rm I \phantom{\dag}} e^{-i\phi}  
        + \sum\limits_{k'k''\sigma} 
        T^{R}_{k'k''\sigma} c^{\rm R\dag}_{k''\sigma} 
        c_{k'\sigma}^{\rm I \phantom{\dag}} e^{-i\phi} e^{i\psi}
        + {\rm h.c.} \ .  
\end{equation} 
Apart from the  microscopic degrees of freedom, (\ref{TUNNEL_HAMILTONIAN})
contains two ``macroscopic''  operators: $e^{\pm i\phi}$ and $e^{\pm i\psi}$.
The first one describes changes of the charge on the transistor island 
due to the tunneling: $e^{i\phi} |N\rangle = |N+1\rangle$. It may be treated 
as an independent degree of freedom if the total
number of electrons on the island is large. We include one more
operator $e^{\pm i\psi}$ which describes the changes of the charge
in the right lead. It acts on $m$, the 
number of electrons which have tunneled through the SET transistor, 
$e^{i\psi} |m\rangle = |m+1\rangle$. 
Since the chemical potential
of the right lead is controlled, $m$ does not appear in any charging 
part of the Hamiltonian.
However,  $e^{\pm i\psi}$ allows us
to keep track of the number of electrons which have passed through
the SET transistor, which is related to the current through the device.

We define the q-bit's Hamiltonian as the part of the total one which 
governs the q-bit's dynamics in equilibrium ($N=0$): 
\begin{equation}
\label{Q-BIT_HAMILTONIAN}
        H_{\rm qb} = E_{\rm qb} (n - Q_{\rm qb})^2 - E_{\rm J}
        \cos \Theta \ .  
\end{equation}
Here $Q_{\rm qb} \equiv -2eV_{n}/2E_{\rm qb}$ is the q-bit's gate
charge, measured in units of $2e$. We concentrate 
on the values of $Q_{\rm qb}$ in an interval around the degeneracy point 
$Q_{\rm qb}=1/2$, so that only the low energy charge  
states $n=0$ and $n=1$ are relevant.    
These states, however, are not appropriate logical 
states of the q-bit since they are not the eigenstates of 
the Hamiltonian (\ref{Q-BIT_HAMILTONIAN}). 
We diagonalize (\ref{Q-BIT_HAMILTONIAN}) in the two charge states
subspace for a fixed value of $Q_{\rm qb}$ 
(which is kept constant between the quantum manipulations
and during the measurement) and denote the corresponding logical
states $|0\rangle$ and $|1\rangle$. In the new basis, up to a constant, 
$H_{\rm qb} = -(1/2)\Delta E\,\sigma_{z}$, where 
$\sigma_z$ is the Pauli matrix and 
\begin{equation}
\label{DELTA_E}
        \Delta E \equiv \sqrt{[E_{\rm qb}(1-2Q_{\rm qb})]^2 + E_{\rm J}^2} \ .
\end{equation}  
The price which we pay for this simplification is that 
the number operator $n$, which appears in the mixed term
of (\ref{CHARGING_ENERGY}), becomes non-diagonal:
\begin{equation}
\label{n_OPERATOR}
        n = {1\over 2} - {1\over 2}\cos\eta\,\sigma_z -
        {1\over 2}\sin\eta\,\sigma_x \ ,
\end{equation} 
with mixing angle $\eta$ given by
$\tan\eta = E_{\rm J} / E_{\rm qb}(1-2Q_{\rm qb})$.
In the quantum regime, which we are considering here, 
$E_{\rm qb} \gg E_{\rm J}$ and,
therefore, one can choose $Q_{\rm qb}$ so that $\tan\eta \ll 1$. 

The interaction Hamiltonian 
(part of the mixed term in (\ref{CHARGING_ENERGY})) 
now becomes
\begin{equation}
\label{INTERACTION_HAMILTONIAN}
        H_{\rm int} = - {1\over 2} E_{\rm int} N 
        (\cos\eta\,\sigma_z + \sin\eta\,\sigma_x) \ ,
\end{equation} 
while the rest of the mixed term ($E_{\rm int} N/2$) as well as all
other remaining terms are collected in the Hamiltonian of the SET transistor:
\begin{equation}
\label{SET_HAMILTONIAN}
        H_{\rm set}= E_{\rm set} (N - Q_{\rm set})^2  
        + H_{\rm L} + H_{\rm R} + H_{\rm I} + H_{\rm T} \ .
\end{equation}
The transistor's gate charge (measured in the units of $e$) became
 $Q_{\rm set} \equiv -(eV_{N} + E_{\rm int}/2) / 2E_{\rm set}$.
The total Hamiltonian reads 
$H = H_{\rm qb} + H_{\rm set} + H_{\rm int}$. One should understand,
however, that the  division chosen is rather arbitrary. The terms 
$H_{\rm qb}$ and $H_{\rm set}$ would not describe the q-bit and
the SET transistor if they were decoupled.

\section{Quantitative description of the measurement, 
time evolution of the reduced density matrix} 

The total system is described by a reduced density matrix 
$\sigma(t) = {\rm Tr}_{\rm L,R,I} \{\rho(t)\}$, where the trace 
is taken over the microscopic states of the leads and of the island. 
In general, the density matrix  $\sigma(i,j;N,N';m,m')$ 
is a matrix in $i,j$, which stand,
for the quantum states of the q-bit 
($|0\rangle$ or $|1\rangle$), in $N$, and in $m$.
However, as has been shown in \cite{Schoeller_PRB} a 
closed set of equations describing the
time evolution of the system can 
be derived where the off-diagonal elements in $N$
have been eliminated. The same is true for
the off-diagonal elements in $m$. Therefore, we need to consider only
the following elements of the density matrix
$\sigma^{N,m}_{i,j} \equiv \sigma(i,j;N,N;m,m)$.
We assume now that at time $t=0$, when the q-bit is prepared in the quantum
state \,$a|0\rangle + b|1\rangle$ as a result of previous 
quantum manipulations, we switch on a transport voltage to the SET transistor.
To proceed we can further reduce the density 
matrix in two different ways to obtain dual descriptions of
the measuring process. 

The first widely used procedure \cite{Gurvitz} is to trace
over $N$ and $m$. This yields a reduced density matrix of
the q-bit $\sigma_{i,j} \equiv \sum_{N,m} \sigma_{i,j}^{N,m}$.
Assuming that at $t=0$ it is in the state
\begin{equation}
\label{SIGMA(0)}
\sigma_{i,j}(0) = 
\left(
\begin{array}{cc}
|a|^2 & a b^* \\
a^* b & |b|^2
\end{array}
\right) \ ,
\end{equation}
the questions are how fast the off-diagonal elements of
$\sigma_{i,j}$ vanish (dephasing), {\sl and} how fast 
the diagonal elements change their original values
(for instance due to transitions induced by the measurement). 
This description 
is enough when one is interested in the quantum properties
of the measured system only (q-bit in our case) and the 
measuring device is used as a source of dephasing 
\cite{Levinson,Aleiner,Buks}. It does not tell us much, however,
about the quantity measured in an experiment, 
namely the current flowing trough
the SET transistor. 

The second procedure is to evaluate the probability distribution of
the number of 
electrons $m$ which have tunneled trough the SET transistor during time $t$:
\begin{equation}
\label{P(m)}
        P(m,t) \equiv \sum_{N,i} \sigma_{i,i}^{N,m}(t) \ .  
\end{equation}
This quantity gives a complete description of the measurement.
At $t=0$ no electrons have tunneled, so $P(m,0) = \delta_{m,0}$.
Then this delta-peak starts to shift in positive $m$
direction and, at the same time, it widens due to shot noise.
Since two states of the q-bit correspond to different 
conductivities (and shift velocities in  $m$ space), 
one may hope that after some time the 
peak splits into two. If after sufficient separation of the two
peaks their weights (integrals) are still close to $|a|^2$
and $|b|^2$, a good quantum measurement has been performed.
Unfortunately, there exist further processes which destroy
this idealized picture. After a long time
the two peaks transform into a broad plateau, since transitions between the 
q-bit's states are induced by the measurement. 
Therefore, one should find an optimum
time for the measurement, so that, on one hand, the two peaks are separate
and, on the other hand, the induced transitions have not yet happened.
In order to describe this  we have to 
analyze the time evolution of the reduced density matrix quantitatively.

\section{Derivation of the master equation}

\begin{figure}  
\epsfysize=6\baselineskip
\centerline{\hbox{\epsffile{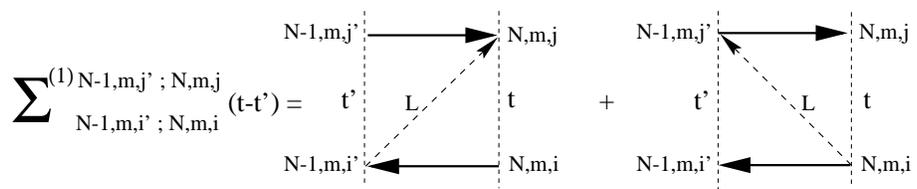}}}
\caption[]{\label{DIAGRAM}
The first order diagram for the transition rates.}
\end{figure}

The Bloch-type or master equations with coherent terms
have only recently been analyzed in the condensed matter physics 
\cite{Nazarov,Gurvitz}. In Ref.~[\onlinecite{Schoeller_PRB}] a diagrammatic
technique has been developed which provides a formally exact
master equation as an expansion in the tunneling strength. 
Only the tunneling term $H_{\rm T}$ is considered a perturbation,
while all other terms constitute the zeroth order Hamiltonian
$H_0 \equiv H - H_{\rm T}$, which is treated exactly. 
The master equation reads
\begin{equation}
\label{MASTER_EQUATION}
        {d\sigma(t)\over dt} - 
        {i\over\hbar}[\sigma(t), H_0] 
        = \int_0^t dt' \Sigma(t-t') \sigma(t') \ ,
\end{equation} 
where the matrix elements of $\Sigma(t',t)$ can be calculated 
diagrammatically using the real-time Keldysh contour
technique \cite{Schoeller_PRB}. The simplest diagram
describing the tunneling through the left junction in first
order perturbation theory (sequential tunneling) is shown
in Fig.~\ref{DIAGRAM}. The dashed lines crossing the diagram 
contribute the following factor to the rate \cite{Schoeller_PRB}:
\begin{equation}
\label{LINE}
        -\,\alpha_{\rm L}\, {({\pi\over\hbar\beta})^2 e^{\pm i\mu_{\rm L}(t-t')} 
        \over \sinh^2\left[{\pi\over\hbar\beta}(t-t' \pm i\delta)\right]} \ , 
\end{equation}
where $\alpha_{\rm L} \equiv \hbar / (4\pi^2 e^2 R_{\rm T_L})$,
$\mu_{\rm L}$ is the electro-chemical potential of the left lead, and
$\delta \equiv 1/\omega_c$ is the inverse frequency 
cut-off. The sign of the $i\delta$ term depends on the 
time-direction of the dashed line \cite{Schoeller_PRB}.
It is minus if the direction of the line with respect 
to the Keldysh contour coincides with its direction 
with respect to the absolute time 
(from left to right), and plus otherwise. For example the right 
part of Fig.~\ref{DIAGRAM} should carry a minus sign, while the left part 
carries a plus sign. For the sign in front of  $i\mu_{\rm L}(t-t')$ 
the rule is as
follows: minus, if the line goes forward with respect to the 
absolute time, and plus otherwise.

For a single SET transistor the horizontal lines correspond 
to trivial exponential 
factors \cite{Schoeller_PRB} $e^{iEt}$. In our case, however,
we have to account for the nontrivial time evolution of the 
q-bit. Therefore  the upper line in the left part
of Fig.~\ref{DIAGRAM} corresponds to 
$\langle N-1,j|e^{-iH_0(t-t')}|N-1,j'\rangle$,
while the lower line corresponds to $\langle N,i|e^{iH_0(t-t')}|N,i'\rangle$.
To calculate these matrix elements 
we diagonalize  
$H_{\rm qb}$ + $H_{\rm int}$ for each value of $N$.
The eigenenergies are 
\begin{equation}
\label{E_01_N}
E_{0,1}^{(N)} = \mp
{1\over 2} \sqrt{(\Delta E + E_{\rm int}N\cos\eta)^2 + 
(E_{\rm int}N\sin\eta)^2} \ ,
\end{equation} 
and the mixing angles $\epsilon_N$ (analogous
to $\eta$) are given by 
$\tan\epsilon_N = 
E_{\rm int} N\sin\eta /(\Delta E + E_{\rm int} N\cos\eta)$.
The matrix elements (propagators) read:
\begin{eqnarray}
\label{MATRIX_ELEMENTS}
&&\langle N,0|e^{-iH_0\Delta t}|N,0\rangle  
=(\cos^2{\epsilon_N\over2}e^{-iE_0^N\Delta t} + 
\sin^2{\epsilon_N\over2}e^{-iE_1^N\Delta t})\  
e^{-iE_{\rm set}^{(N)}\Delta t} \ , \nonumber \\
&&\langle N,1|e^{-iH_0\Delta t}|N,1\rangle  
=(\cos^2{\epsilon_N\over2}e^{-iE_1^N\Delta t} + 
\sin^2{\epsilon_N\over2}e^{-iE_0^N\Delta t})\ 
e^{-iE_{\rm set}^{(N)}\Delta t} \ , \nonumber \\
&&\langle N,1|e^{-iH_0\Delta t}|N,0\rangle  
={1\over2}\sin\epsilon_N (e^{-iE_0^N\Delta t}-e^{-iE_1^N\Delta t})\ 
e^{-iE_{\rm set}^{(N)}\Delta t} \ ,
\end{eqnarray}
where $E_{\rm set}^{(N)} \equiv E_{\rm set}(N-Q_{\rm set})^2$.

We analyze now the rates in Fig.~\ref{DIAGRAM} for
different choices of q-bit's indices in the regime 
$\Delta E \gg E_{\rm int}, E_{\rm J}$.
There the mixing angles are small, 
$\epsilon_N \propto N E_{\rm int} E_{\rm J} /(\Delta E)^2$,
for all relevant values of $N$. Hence, we 
keep only terms linear in $\epsilon_N$.  
The simplest transition 
$(i'=0,j'=0,N-1,m) \rightarrow (i=0,j=0,N,m)$ is described by 
\begin{equation}
\label{00to00RATE}
        \Sigma^{(1)\ N-1,m,0;N,m,0}_{\ \ \ \ N-1,m,0;N,m,0}(\Delta t) =
        {-\alpha_{\rm L}({\pi\over\hbar\beta})^2 e^{-i\tilde E\Delta t}  
        \over \sinh^2\left[{\pi\over\hbar\beta}(\Delta t + i\delta)\right]} + 
        {\rm c.c.} \ ,
\end{equation}  
where $\tilde E$ stands here for 
$\mu_{\rm L} + (E_{\rm set}^{(N-1)} - E_{\rm set}^{(N)}) + 
(E_0^{N-1} - E_0^{N})$.

The form of the master equation (\ref{MASTER_EQUATION}) suggests 
the use of a Laplace transformation, after which the last
term in (\ref{MASTER_EQUATION}) becomes $\Sigma(s)\sigma(s)$. 
We Laplace transform (\ref{00to00RATE}) in the regime 
$s \ll \tilde E$, i.e.\ we assume the density matrix $\sigma$
to change slowly on the time scale given by $\hbar/\tilde E$.
This assumption should be verified later for self-consistency.
At zero temperature ($\beta \rightarrow \infty$) and for
$\delta \rightarrow 0$ we obtain:
\begin{eqnarray}
\label{00to00LAPLACE}
    &&\Sigma^{(1)\ N-1,m,0;N,m,0}_{\ \ \ \ N-1,m,0;N,m,0}(s)
    =2\alpha_{\rm L} {\rm Re} \left[ (s+i\tilde E) e^{i\delta(s+i\tilde E)}
    E_1[i\delta(s+i\tilde E)] \right]  
\nonumber \\
    &&\approx 2\pi\alpha_{\rm L} \tilde E \Theta(\tilde E) -
    2\alpha_{\rm L} s(1+\gamma + \ln(|\delta\tilde E|)) \ ,
\end{eqnarray}
where $E_1[...]$ is the exponential integral 
and $\gamma \approx 0.58$ is Euler's constant.
Denoting the diverging factor 
$[1+\gamma + \ln(|\delta\tilde E|)]$
by $D(\tilde E)$ and performing the 
inverse Laplace transform we arrive at:
\begin{eqnarray}
\label{00to00SLOW}
        \Sigma^{(1)\ N-1,m,0;N,m,0}_{\ \ \ \ N-1,m,0;N,m,0}(\Delta t) 
        \approx 2\pi\alpha_{\rm L} 
        \tilde E \Theta(\tilde E)\delta(\Delta t+0) -
        2\alpha_{\rm L} D(\tilde E) \delta'(\Delta t + 0) \ .
\end{eqnarray}
(Note that (\ref{00to00SLOW}) is equivalent to 
(\ref{00to00RATE}) only as a kernel in the convolution
(\ref{MASTER_EQUATION}) when applied to slowly changing
matrix elements of $\sigma$). The first term of 
(\ref{00to00SLOW}) is the usual Golden Rule tunneling
rate corrected with respect to the additional
charging energy corresponding to the quantum
state $|0\rangle$ of the q-bit, $E_0^{N-1}-E_0^{N}$.
The second (diverging) part of (\ref{00to00SLOW}) produces a term 
proportional to ${d\over dt}\,\sigma_{0,0}^{N-1,m}$. 
One can take this term to the LHS of (\ref{MASTER_EQUATION}) so that
the time derivative in the LHS will look like 
${d\over dt}[\sigma_{0,0}^{N,m} 
- 2\alpha_{\rm L} D(\tilde E)\sigma_{0,0}^{N-1,m}]$.
We analyze all possible choices of the q-bit's indices
in Fig.~\ref{DIAGRAM} and arrive at the conclusion that
the diverging terms have always the same structure as the 
coherent terms in the LHS of (\ref{MASTER_EQUATION}).
Moreover, if we neglect some energy corrections of
order $E_{\rm int}$, we may
incorporate all of these terms to the LHS of 
(\ref{MASTER_EQUATION}), so that 
the master equation reads:
\begin{equation}
\label{MASTER_EQUATION_CORRECTED}
        (1+\alpha_{\rm L} A+\alpha_{\rm R} B) \left[
        {d\sigma(t)\over dt} - {i\over\hbar}[\sigma(t), H_0] 
        \right] = \Gamma \sigma(t) \ ,
\end{equation}
where $A$ and $B$ are three-diagonal matrices in the $N$ and $m$
spaces, composed of the diverging factors of the type of $D(\tilde E)$,
while $\Gamma$ is the regular local part of $\Sigma(t-t')$.
 
We expect that without the approximation of energies in the
diverging terms the structure of (\ref{MASTER_EQUATION_CORRECTED})
would be the same, with $A$ and $B$ being more complicated matrices,
which would include some mixing in the space of the q-bit's states. 
Finally, we note that for any physically reasonable choice of the cut-off 
$\delta$, the logarithmically divergent factors in the matrices $A$ and $B$
are of order one, and, therefore, the mixing
corrections to the unit matrix in the LHS of 
(\ref{MASTER_EQUATION_CORRECTED}) are small. We multiply the 
master equation (\ref{MASTER_EQUATION_CORRECTED}) 
by $(1+\alpha_{\rm L} A+\alpha_{\rm R} B)^{-1} 
\approx (1-\alpha_{\rm L} A-\alpha_{\rm R} B)$ 
from the left, so that the mixing 
corrections move to the RHS.
Since $\Gamma$ is linear in $\alpha_{\rm L}$
and $\alpha_{\rm R}$, the mixing corrections are quadratic. We drop them
in the framework of the first order perturbation theory.
The master equation to be analyzed thus becomes:
\begin{equation}
\label{MASTER_EQUATION_FINAL}
        {d\sigma(t)\over dt} - {i\over\hbar}[\sigma(t), H_0]= 
        \Gamma \sigma(t) \ .
\end{equation}

If the applied voltage is not too high (the exact criterion to be
specified) we may consider only two charge states of the SET transistor,
$N=0,1$. We perform a Fourier transform in $m$ space 
$\sigma^{N}_{i,j}(k) \equiv \sum_m \sigma^{N,m}_{i,j}e^{ikm}$.
To shorten formulas we introduce $A^N \equiv \sigma_{0,0}^{N}(k)$,
$B^N \equiv \sigma_{1,1}^{N}(k)$, 
$C^N \equiv \sum\limits_m {\rm Re}\,\sigma_{0,1}^{N,m} \,e^{ikm}$,
and $D^N \equiv \sum\limits_m {\rm Im}\,\sigma_{0,1}^{N,m} \,e^{ikm}$. 
This enables us to rewrite (\ref{MASTER_EQUATION_FINAL}) as:
\begin{eqnarray}
\label{ME1}
\dot A^0 &=&
-\Gamma_{\rm L_0} A^0 + \Gamma_{\rm R_0} e^{ik} A^1 
-\omega_{\rm L} C^0
-\omega_{\rm R} e^{ik} C^1 \\
\label{ME2}
\dot A^1 &=&
\Gamma_{\rm L_0} A^0  -
\Gamma_{\rm R_0} A^1 - \Omega D^1 
+\omega_{\rm L} C^0
+\omega_{\rm R} C^1 \\
\label{ME3}
\dot B^0 &=&
-\Gamma_{\rm L_1} B^0 + \Gamma_{\rm R_1} e^{ik} B^1 
-\omega_{\rm L} C^0
-\omega_{\rm R} e^{ik} C^1 \\
\label{ME4}
\dot B^1 &=&
\Gamma_{\rm L_1} B^0 -
\Gamma_{\rm R_1} B^1 + \Omega D^1 
+\omega_{\rm L} C^0
+\omega_{\rm R} C^1 \\
\label{ME5}
\dot C^0 &=&
-\Delta E^{0} D^0 - \Gamma_L C^0 + \Gamma_R e^{ik}C^1 
-{\omega_{\rm L}\over 2}(A^0-B^0)-{\omega_{\rm R}\over 2}e^{ik}(A^1-B^1) \\ 
\label{ME6}
\dot C^1 &=&
-\Delta E^{1} D^1 + \Gamma_L C^0 - \Gamma_R C^1 
+{\omega_{\rm L}\over 2}(A^0-B^0)+{\omega_{\rm R}\over 2}(A^1-B^1) \\
\label{ME7}
\dot D^0 &=& 
\Delta E^{0} C^0 -\Gamma_L D^0 + \Gamma_R e^{ik} D^1 \\
\label{ME8}
\dot D^1 &=& 
\Delta E^{1} C^1 +\Gamma_L D^0 - \Gamma_R D^1 
+{\Omega\over 2}(A^1-B^1) \ .
\end{eqnarray}
Here $\Delta E^{0,1} \equiv E_1^{0,1} - E_0^{0,1}$ are the 
energy differences between the q-bit's states for $N=0$ and
$N=1$ respectively, and $\Omega \equiv E_{\rm int}\sin\eta$ 
is the coefficient in the mixing term in $H_{\rm int}$
for $N=1$ (see (\ref{INTERACTION_HAMILTONIAN})). 
The terms proportional to $\Delta E^{0,1}$ and $\Omega$
originate from the coherent part of (\ref{MASTER_EQUATION_FINAL}).
The tunneling rates which appear in the four last equations for the 
off-diagonal elements (\ref{ME5}-\ref{ME8}) are given by
\begin{eqnarray}
\label{ORIGINAL_RATES}
        \Gamma_L &\equiv& 2\pi\alpha_{\rm L}
        [\mu_{\rm L} - E_{\rm set}(1-2Q_{\rm set})] 
\nonumber \\  
        \Gamma_R &\equiv& 2\pi\alpha_{\rm R}
        [-\mu_{\rm R} + E_{\rm set}(1-2Q_{\rm set})]
\end{eqnarray}
The rates 
\begin{eqnarray}
\label{CORRECTED_RATES}
        \Gamma_{\rm L_{0,1}} \equiv \Gamma_L \pm \Delta\Gamma_L \nonumber \\
        \Gamma_{\rm R_{0,1}} \equiv \Gamma_R \pm \Delta\Gamma_R
\end{eqnarray}
appearing in the equations for the diagonal 
elements (\ref{ME1}-\ref{ME4}) are corrected due to the 
charging energy induced by the q-bit:
\begin{eqnarray}
\label{DELTA_GAMMA}
        \Delta\Gamma_{\rm L} &\equiv& 2\pi\alpha_{\rm L}(E_0^0 - E_0^1) 
        \nonumber \\
        \Delta\Gamma_{\rm R} &\equiv& -2\pi\alpha_{\rm R}(E_0^0 - E_0^1)  
        \ .
\end{eqnarray}
These correction are, actually, responsible for the separation 
of the peaks. In the regime $\tan\eta \ll 1$, which we assume here,
$|\Delta\Gamma_{\rm L,R}| \approx 2\pi\alpha_{\rm L,R} E_{\rm int}$.
The rest are small mixing terms, 
$\omega_{\rm L,R} \equiv \pi\alpha_{\rm L,R}\sin\epsilon_1\Delta E^1$,
which also originate from diagrams of the type in 
Fig.~\ref{DIAGRAM}. Note, that we assume that only the two rates
given in (\ref{ORIGINAL_RATES}) are nonzero (two-state approximation).
Moreover we assume that the q-bit's charging energy corrections 
can at most change these two rates, but they can not switch on  
any other rate or switch off one of the two in (\ref{ORIGINAL_RATES}).

\section{Qualitative analysis of the master equation}
First, we analyze the system (\ref{ME1}-\ref{ME8}) qualitatively.  
Imagine that we can ``switch off'' the Josephson
coupling during the measurement. Then all
the mixing terms in (\ref{ME1}-\ref{ME8}), i.\,e.\, 
those proportional to $\Omega$ and $\omega_{\rm L,R}$ disappear,
and the system factorizes into three independent groups. 
The first one (\ref{ME1},\ref{ME2}),
\begin{eqnarray}
\label{A_SUBSYSTEM}
        \dot A^0 &=&
        -\Gamma_{\rm L_0} A^0 + \Gamma_{\rm R_0} e^{ik} A^1 \nonumber \\
        \dot A^1 &=&
        \Gamma_{\rm L_0} A^0  -
        \Gamma_{\rm R_0} A^1 \ ,
\end{eqnarray}
has plain wave solutions ($\propto e^{i\omega t}$).
The standard analysis gives for eigenvalues:
\begin{equation}
\label{A_EIGENVALUES}
        \omega_{1,2} = {i\over 2}(\Gamma_{\rm L_0}+\Gamma_{\rm R_0})
        \left\{1 \pm 
        \left[1+{4\Gamma_{0}(e^{ik}-1)
        \over \Gamma_{\rm L_0}+\Gamma_{\rm R_0}}\right]^{1\over 2}\right\} \ ,
\end{equation}
where 
\begin{equation}
\label{GAMMA_ZERO}
\Gamma_{0} \equiv {\Gamma_{\rm L_0}\Gamma_{\rm R_0}
\over \Gamma_{\rm L_0}+\Gamma_{\rm R_0}}
\end{equation} 
is the total transport rate corresponding to the q-bit is 
in the state $|0\rangle$.
When $k$ is small $\omega_1 \approx i(\Gamma_{\rm L_0}+\Gamma_{\rm R_0})$,
while $\omega_2 \approx \Gamma_{\rm 0}k + 
i\Gamma_{0}(1+{2\Gamma_{0} \over \Gamma_{\rm L_0}+\Gamma_{\rm R_0}})k^2$.
Since $\omega_1$ is a large imaginary number, already after a short
time, $1/|\omega_1|$, only the second eigenvector 
($A^{1}/A^{0} = \Gamma_{\rm L_0}/\Gamma_{\rm R_0}$)
survives. This eigenvector multiplies  
a wave packet propagating with the group velocity $\Gamma_{0}$.
The wave packet widens due to shot noise of the single electron
tunneling, it's width being 
given by $\sqrt{\Gamma_{0} t}$ (the second
imaginary term in the expression for $\omega_2$).

Analogously the second group of equations (\ref{ME3},\ref{ME4})
gives a wave packet with the group velocity 
$\Gamma_{1} \equiv {\Gamma_{\rm L_1}\Gamma_{\rm R_1}
\over (\Gamma_{\rm L_1}+\Gamma_{\rm R_1})}$ and the width 
$\approx \sqrt{\Gamma_{1} t}$. The two peaks correspond
to the q-bit in the states $|0\rangle$ and $|1\rangle$, respectively.
They separate when their width is smaller then
the distance between their centers 
$\sqrt{\Gamma_{0}t}+\sqrt{\Gamma_{1}t} \le |\Gamma_{0}-\Gamma_{1}|t$.  
After this time, 
\begin{equation}
\label{MEASUREMENT_TIME}
        t_{\rm ms} \equiv |\sqrt{\Gamma_{0}}-\sqrt{\Gamma_{1}}|^{-2} \ ,
\end{equation} 
which we denote as the measurement time, the process
can constitute a quantum measurement.
Similar expressions have been obtained in 
Refs.~[\onlinecite{Levinson,Aleiner,Gurvitz}], 
where they have been denoted as dephasing time.

To get a clue for the dephasing we analyze the third 
group of equations (\ref{ME5}-\ref{ME8}) at $k=0$
(the trace over $m$ is equivalent to $k=0$).
These equations may be combined into two complex ones:
\begin{eqnarray}
\label{CD_SUBSYSTEM}
\dot\sigma_{0,1}^{0} = i\Delta E^{0}\sigma_{0,1}^{0}
-\Gamma_L\sigma_{0,1}^{0} + \Gamma_R\sigma_{0,1}^{1}
\nonumber \\
\dot\sigma_{0,1}^{1} = i\Delta E^{1}\sigma_{0,1}^{1}
+\Gamma_L\sigma_{0,1}^{0} - \Gamma_R\sigma_{0,1}^{1}
\end{eqnarray}
The standard analysis shows that if 
$dE \equiv |\Delta E^{1} - \Delta E^{0}| \approx E_{\rm int} 
\ll (\Gamma_L + \Gamma_R)$
the imaginary parts of the eigenvalues are 
${\rm Im}\,\omega_1 \approx (\Gamma_L + \Gamma_R)$ and 
${\rm Im}\,\omega_2 \approx {dE^2 \over 4 (\Gamma_L + \Gamma_R)}$.
In the opposite limit 
$dE \gg (\Gamma_L + \Gamma_R)$ the imaginary parts are
${\rm Im}\,\omega_1 \approx \Gamma_L$ and 
${\rm Im}\,\omega_2 \approx \Gamma_R$. The first limit is physically
more relevant (we have assumed parameters in this regime),
although the second one is also possible if the tunneling
is too weak or the coupling between the q-bit ant the SET transistor is
too strong. In both limits the dephasing time, which is defined as the
the longer of the two times,
\begin{equation}
\label{TAU_PHI_DEFINITION}
\tau_{\phi} \equiv {\rm max}\{[{\rm Im}\,\omega_1]^{-1},
[{\rm Im}\,\omega_2]^{-1}\}
\end{equation}
is parametrically
different from the measurement time (\ref{MEASUREMENT_TIME}).
In the first limit, $dE \ll (\Gamma_L + \Gamma_R)$, it is
\begin{equation}
\label{DEPHASING_TIME}
\tau_{\phi} = {4(\Gamma_L + \Gamma_R)\over dE^2} \propto \alpha_{\rm L,R} \,
\end{equation}
while $t_{\rm ms} \propto \alpha_{\rm L,R}^{-1}$. 
One can check that in the whole range of validity
of our approach the measurement time exceeds the dephasing time, 
$t_{\rm ms} > \tau_\phi$. This is consistent
with the fact that a ``good'' quantum measurement should completely
dephase a quantum state. In Refs.~[\onlinecite{Levinson,Aleiner,Gurvitz}],
where different systems have been discussed,
the expressions for the resulting dephasing time were given by 
expressions similar to 
(\ref{MEASUREMENT_TIME}), thus $\tau_\phi = t_{\rm ms}$.

In our example the dephasing time is shorter than the measurement time. 
The reason for this is, probably, the presence of the additional 
uncontrolled environment provided by the middle island of the SET transistor. 
The transport of electrons occurs via a real state of the island $N=1$.
In different transitions the island may be left in different 
microscopic states, even though the same number of electrons have passed. 
To put it in the 
language of Ref.~[\onlinecite{Stern_Aharonov_Imry}], the initial state 
of the system $(a|0\rangle + b|1\rangle)\,|\chi\rangle\,|m=0\rangle$
evolves into $a|0\rangle\,|\chi_0\rangle\,|m_0\rangle
+b|1\rangle\,|\chi_1\rangle\,|m_1\rangle$, where $|\chi\rangle$ stands
for the quantum state of the uncontrolled environment. One may imagine
a situation when $m_0 = m_1$, but $|\chi_0\rangle$ and $|\chi_1\rangle$
are orthogonal. In this situation the dephasing has occured but no
measurement has been performed. 

The additional environment plays, actually, a positive role,
i.\,e. it helps us to perform a quantum measurement, provided it 
dephases the state of the q-bit only when the system is driven out of 
equilibrium. This is because the dephasing suppresses the transitions 
between the states of the q-bit (Zeno effect).     

\section{The mixing time}
Finally, we analyze what happens if we take into account the 
mixing terms in the system (\ref{ME1}-\ref{ME8}). We assume 
$k=0$ and investigate the eigenvalues of the eight by 
eight matrix formed by the coefficients of (\ref{ME1}-\ref{ME8}).
Note that in the discussion above we have calculated
all the eight eigenvalues for $E_{\rm J} = 0$ (the two eigenvalues
of the complex system (\ref{CD_SUBSYSTEM}) are doubled when
one considers it as a system of four real equations). 
In the diagonal part there were two zeros, 
which corresponded to two conserved
quantities, $A^{0} + A^{1}=\sigma_{0,0}$ and 
$B^{0} + B^{1}=\sigma_{1,1}$. Six other eigenvalues were
large compared to the amplitudes of the mixing terms.
It is clear, that switching on the mixing
changes only slightly the values of the six large 
eigenvalues. Moreover, one of the eigenvalues is always zero. 
This corresponds to the conservation
of the total trace $A^{0}+A^{1}+B^{0}+B^{1}=1$. The last (8th)
eigenvalue acquires now a small imaginary part and this
gives the time scale of the mixing between the two states
of the q-bit.

We do not have an analytical expression for the mixing time,
but we can estimate it for a concrete physical situation.
At the degeneracy point,
we have $\Gamma_{\rm L}=\Gamma_{\rm R}$, and the corrections
to the rates (\ref{DELTA_GAMMA}) cancel each other, thus,
 no measurement is performed.
Therefore, we choose $Q_{\rm set}$ far enough from the 
degeneracy point, which is $Q_{\rm set}=1/2$, so that 
$\Gamma_{\rm L} < \Gamma_{\rm R}$ and the Coulomb blockade energy
$E_{\rm CB}\equiv E_{\rm set}(1-2Q_{\rm set})$ 
is of the order of $E_{\rm set}$.  
To satisfy the conditions for the Golden Rule 
(see (\ref{00to00RATE}) and the 
discussion thereafter) we assume the chemical potential of 
the left lead $\mu_{\rm L}=V/2$ to exceed the Coulomb blockade 
energy by an amount of the order of $E_{\rm CB} \propto E_{\rm set}$
and assume $E_{\rm CB}$ to be the largest energy scale of the system:
$E_{\rm CB} \gg \Delta E$. The transport voltage should not, however,
exceed the limit, after which the third charge state of the SET
transistor $N=-1$ becomes involved. Thus 
$V/2 < E_{\rm set}(1+2Q_{\rm set})$ and $Q_{\rm set}$
should be chosen far enough from zero as well. 
In this regime we estimate the mixing time as   
\begin{equation}
\label{MIXING_TIME}
t_{\rm mix}^{-1} 
\propto 2\pi\alpha {E_{\rm int}^2 E_{\rm J}^2 \over (\Delta E)^4}
E_{\rm set} \ ,
\end{equation}
where $\alpha_{\rm L} = \alpha_{\rm R} \equiv \alpha$.           
The measurement time in the same regime is given approximately by
\begin{equation}
\label{MEASUREMENT_TIME_APPROX}
t_{\rm ms}^{-1} \propto 
2\pi\alpha {E_{\rm int}^2 \over E_{\rm set}} \ .
\end{equation} 
The exact values of $Q_{\rm set}$ and $V$ would
determine the numerical coefficients in front of 
(\ref{MIXING_TIME}) and (\ref{MEASUREMENT_TIME_APPROX}). 
Thus, $t_{\rm ms}/ t_{\rm mix} \propto
E_{\rm J}^2 E_{\rm set}^2 / (\Delta E)^4$. One recognizes 
two competing ratios here: $E_{\rm J} / \Delta E$, which 
is small, and $E_{\rm set} / \Delta E$, which is large.
The condition ${t_{\rm ms}/t_{\rm mix}} \ll 1$, thus, imposes an 
additional restriction on the parameters of the system.

\section{Discussion}
\begin{figure}  
\epsfysize=40\baselineskip
\centerline{\hbox{\epsffile{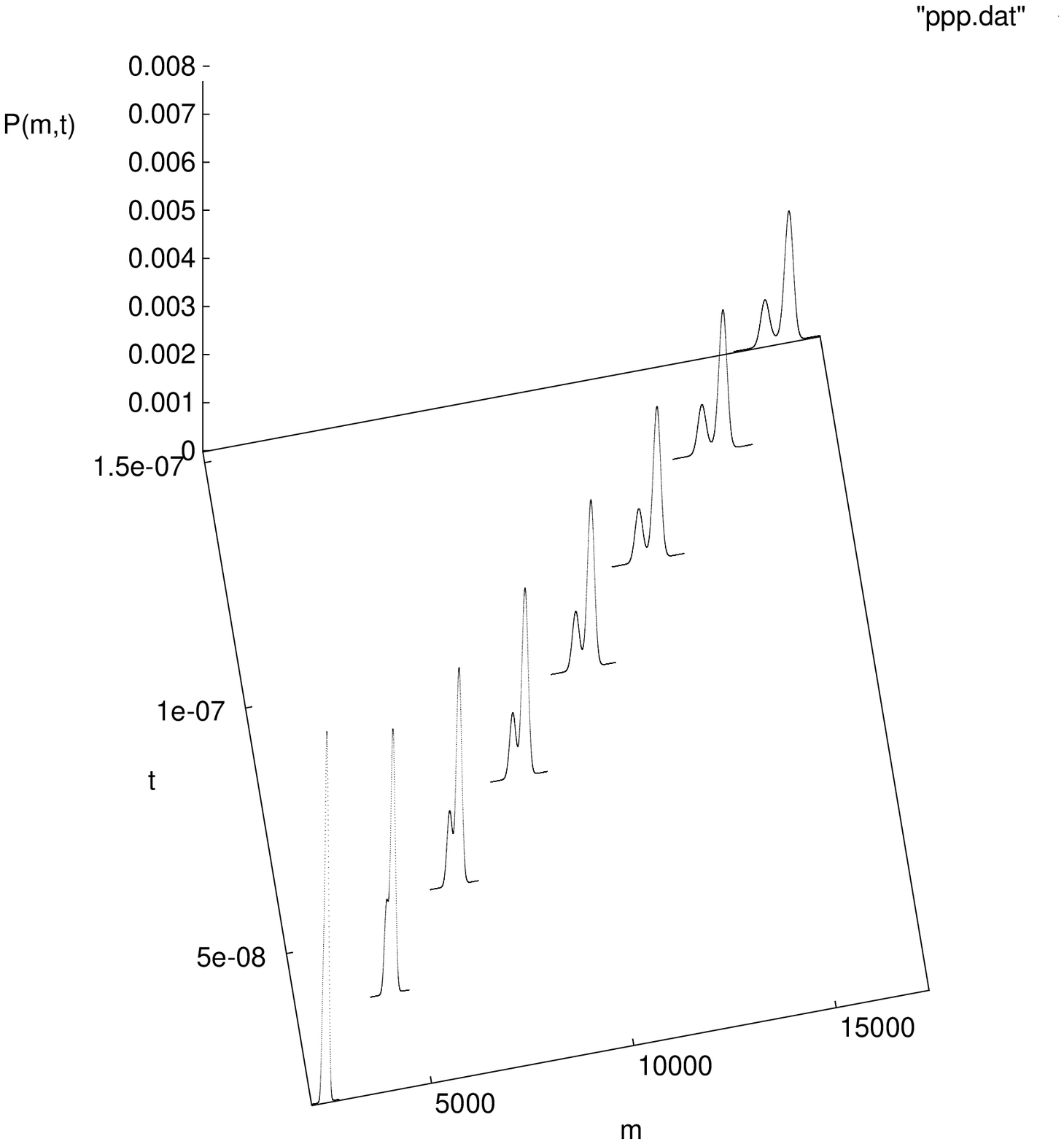}}}
\caption[]{\label{PLOT0.1}
$P(m,t)$, the probability that $m$ electrons have tunneled during time
$t$. The parameters are those given in the text, $E_{\rm J} = 0.1$K. 
The time is measured in seconds.
The initial amplitudes of the q-bit's states: $a = \sqrt{0.75}$, 
$b=\sqrt{0.25}$.} 
\end{figure}

\begin{figure}
\epsfysize=40\baselineskip
\centerline{\hbox{\epsffile{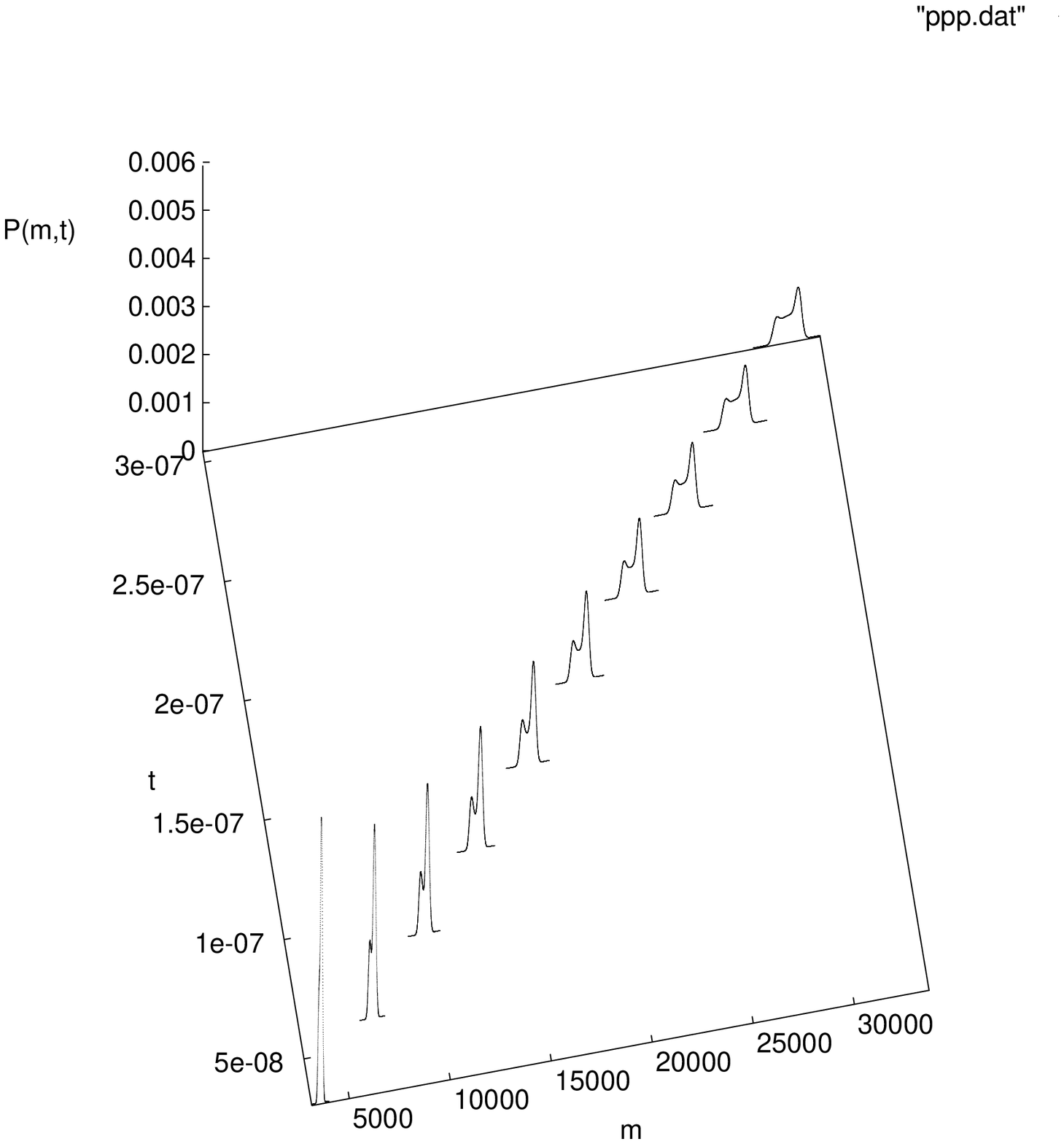}}}
\caption[]{\label{PLOT0.3}
$P(m,t)$, the probability that $m$ electrons have tunneled during time
$t$. The parameters are those given in the text, $E_{\rm J} = 0.25$K.
The time is measured in seconds.
The initial amplitudes of the q-bit's states: $a = \sqrt{0.75}$,
$b=\sqrt{0.25}$.} 
\end{figure}

To show that all the conditions assumed in this paper are 
realistic we calculate the charging energies $E_{\rm set}$,
$E_{\rm qb}$ and $E_{\rm int}$ for the following case: the capacitance
of the Josephson junction $C_{\rm J} = 2.0\times 10^{-16}$F,
the capacitances of the normal junctions 
$C_{\rm N}=1.0\times 10^{-17}$F and the capacitances of 
all other capacitors $C = 2.5\times 10^{-18}$F. 
We obtain: $E_{\rm set}\approx 20$K, $E_{\rm qb}\approx 10$K, 
$E_{\rm int}\approx 0.5$K. 
Taking $Q_{\rm qb}=0.35$, $Q_{\rm set} = 0.15$ and $eV = 48$K 
we get $\Delta E \approx 3$K, $E_{\rm CB} \approx 14$K, and 
$V/2 - E_{\rm CB} \approx 10$K. We also assume $2\pi\alpha = 0.1$.
The measurement time in this regime
is $t_{\rm ms} \approx 0.38 \times 10^{4} \hbar/(k_{\rm B}\,1{\rm K}) \approx
0.28 \times 10^{-7}$s.
For this choice of parameters we calculate $t_{\rm mix}$
numerically, assuming first $E_{\rm J} = 0.1$K, and we obtain 
$t_{\rm mix} \approx 1.4 \times 10^{5} \hbar/(k_{\rm B}\,1{\rm K}) \approx 
1.0 \times 10^{-6}$s. Thus $t_{\rm mix}/t_{\rm ms} \approx 35$ and 
the separation of peaks should occur much earlier than the transitions
happen. Indeed, the numerical simulation of the  
system (\ref{ME1}-\ref{ME8}) for those parameters given 
above shows almost ideal separation of peaks (see Fig.~\ref{PLOT0.1}).
Then, we calculate $t_{\rm mix}$ for $E_{\rm J} = 0.25$K, 
and we obtain $t_{\rm mix}/t_{\rm ms} \approx 6$. This is a marginal
situation. The numerical simulation in this case shows 
(see Fig.~\ref{PLOT0.3}) that the peaks, first, start to 
separate, but, later, the valley between the peaks is filled due to 
the transitions. 

\begin{figure}  
\epsfysize=20\baselineskip
\centerline{\hbox{\epsffile{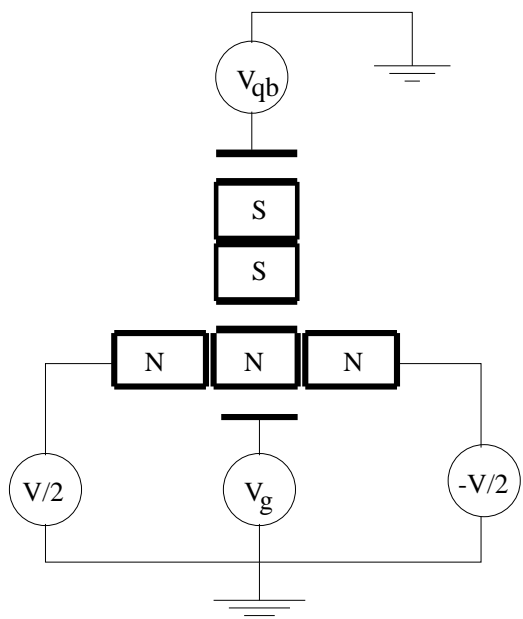}}}
\caption[]{\label{CIRCUIT}
The circuit consisting of a q-bit plus a SET transistor
used as a measuring device.}
\end{figure}

In this paper we have demonstrated  that the current through a
single-electron transistor can serve as a measurement of the
quantum state of the q-bit, in the sense that in the case of a
superposition of two eigenstates it gives one or the other result
with the appropriate probabilities. This should be distinguished from
another question, namely whether it is possible to demonstrate that an
eigenstate of a q-bit can actually be a superposition of two different
charge states, i.e.\ whether it depends on the
mixing angle $\eta$  as described by Eq.\ (\ref{n_OPERATOR}). This question
has been addressed in the experiments of 
Refs.~[\onlinecite{Bouchiat,Bouchiat_PhD}].
They used a setup similar to the one shown in Fig.~\ref{CIRCUIT}, a 
single-Cooper-pair box coupled to a single-electron transistor. They could
demonstrate that the expectation, i.e.\ the average value of the
charge in the box varies continuously as a function of the applied
gate voltage as follows from (\ref{n_OPERATOR}). 

Our theory can also describe the type of measurements performed in 
Refs.~[\onlinecite{Bouchiat,Bouchiat_PhD}]. For this purpose 
we analyze the rates in the 
master equation (\ref{MASTER_EQUATION_FINAL})
for general values of the mixing angle $\eta$,
relaxing the requirement $\tan\eta \ll 1$.
Then, for our approach to be valid, we must have 
$E_{\rm int} \ll E_{\rm J}$, so that $\tan\epsilon_N \ll 1$.
In this regime each eigenstate of the q-bit,
$|0\rangle$ or $|1\rangle$,  corresponds to a single,
though $\eta$-dependent propagation velocity ($\Gamma_0$ or $\Gamma_1$).
Thus, if the q-bit is prepared in one of its eigenstates,
then even at the degeneracy point 
($\eta = \pi/2$) where the eigenstates are equally weighted
superpositions of two charge states, one would observe only
one peak. 
We have calculated $\Gamma_0$ as a function of $\eta$ using 
(\ref{ORIGINAL_RATES}), 
(\ref{CORRECTED_RATES}), (\ref{DELTA_GAMMA}), and
(\ref{GAMMA_ZERO}) and obtained curves (not shown here)
very similar to those in the experiments. 
It should be added that near degeneracy our setup would not be
efficient in projecting onto the eigenstates anymore,
since the difference between the velocities of the peaks, 
$|\Gamma_0 - \Gamma_1|$, vanishes near the degeneracy point.  

To conclude we have shown that a single-electron transistor
capacitively coupled to a q-bit may serve as a quantum measuring device
in an accessible range of parameters. We have described the process of 
measurement by deriving the time evolution of the reduced density matrix 
and we have discussed two dual ways to further reduce it. 
One way, in which the density 
matrix of the q-bit is obtained, provides the dephasing time,
while the other, in which the number of tunneled electrons is
counted, provides the time of measurement. We have shown that, 
in our case, the dephasing time was shorter than the measurement 
time, and we have discussed the physical meaning of this result. 
Finally, we have estimated the mixing time, i.\,e. 
the time scale on which the transitions induced by the measurement occur. 
We have shown that it may be made longer than the measurement time
with current technology. 
      
\acknowledgments
We thank J. K\"onig, Y.~Gefen, Y.~Levinson, 
J.~E.~Mooij, T.~Pohjola, H.~Schoeller, E.~Ben-Jacob, 
C.~Bruder, Y.~Makhlin, Z.~Hermon for stimulating
discussions. This work is supported by the Graduiertenkolleg 
``Kollektive Ph\"anomene im Festk\"orper'', by
the SFB 195 of the DFG, and by the German Israeli Foundation 
(Contract G-464-247.07/95).

\bibliographystyle{unsrt}
\bibliography{ref}

\vskip 3cm

\newpage

\newpage

\end{document}